\documentclass[12pt]{article}
\usepackage{amsmath}
\usepackage{amssymb}
\usepackage{epsfig}
\usepackage{euscript}
\usepackage{fancybox}
\usepackage{color}
\usepackage{graphicx}

\def\mathswitchr#1{\relax\ifmmode{\mathrm{#1}}\else$\mathrm{#1}$\fi}

\newcommand {\pslash}{\hbox{$\not\hbox{\kern-2.3pt $p$}$}}

\usepackage{cite}

\usepackage{epic}

\def\alf1{ {\alpha\over\pi} }

\begin{document}
\begin{titlepage}
\begin{flushright}
{\bf BU-HEPP-08-02 }\\
{\bf Mar., 2008}\\
\end{flushright}
 
\begin{center}
{\Large Review of Applications of YFS-Style Resummation in 
Quantum Field Theory via Monte Carlo Methods$^{\dagger}$}
\end{center}

\vspace{2mm}
\begin{center}
{\bf   B.F.L. Ward}\\
\vspace{2mm}
{\em Department of Physics,\\
 Baylor University, Waco, Texas, 76798-7316, USA}\\
\end{center}

\vspace{5mm}
\begin{center}
{\bf   Abstract}
\end{center}
We review the application of exact, amplitude-based, YFS-style resummation in
quantum field theory via Monte Carlo methods. 
\\
\vskip 20mm
\begin{center}
{Invited talk presented at the 2008 Cracow Epiphany Conference\\
in honor of the $60^{th}$ birthday of Prof. S. Jadach}
\end{center}
\vspace{10mm}
\renewcommand{\baselinestretch}{0.1}
\footnoterule
\noindent
{\footnotesize
\begin{itemize}
\item[${\dagger}$]
Work partly supported by US DOE grant DE-FG02-05ER41399 and 
by NATO Grant PST.CLG.980342.
\end{itemize}
}

\end{titlepage}

\def\Kmax{K_{\rm max}}\def\ieps{{i\epsilon}}\def\rQCD{{\rm QCD}}
\renewcommand{\theequation}{\arabic{equation}}
\font\fortssbx=cmssbx10 scaled \magstep2
\renewcommand\thepage{}
\parskip.1truein\parindent=20pt\pagenumbering{arabic}\par
\section{\bf Preface}
It is with great pleasure that I present this review of the application
of YFS-style~\cite{yfs} exact, amplitude based 
resummation via Monte Carlo methods
on the occasion of the $60^{th}$ birthday of Prof. S. Jadach, my friend
and collaborator since 1985. In the review, we intend to highlight some of 
the many pioneering contributions
which Prof. Jadach has made to this important subject. We are all
grateful to him for all that he has taught us about the subject.
\section{\bf Introduction}\label{intro}\par
The theoretical foundation of the subject of this discussion
is the pioneering paper by D.R. Yennie, S.C. Frautschi and H. Suura
published already in 1961~\cite{yfs}. In this paper, the exact result
for the processes $f_1(p_1)+f_2(p_2)\rightarrow f_3(p_3)+f_4(p_4)+n(\gamma)$ is given as
\begin{equation}
\begin{split}
d\sigma_{\rm exp}&=e^{\rm 2\alpha\Re{B}+2\alpha\tilde{B}}\sum_{n=0}^\infty\frac{1}{n!}\int\prod_{j=1}^n{d^3
k_j\over k_j}\int{d^4y\over(2\pi)^4}e^{iy\cdot(p_1+p_2-p_3-p_4-\sum k_j)+
D}\\
&*\bar\beta_n(k_1,\ldots,k_n){d^3p_3\over p_3^{\,0}}{d^3p_4\over
p_4^{\,0}}
\end{split}
\label{subp15}
\end{equation}
where the hard photon residuals, 
$\tilde{\bar\beta}_n(k_1,\ldots,k_n)$, as defined in Ref.~\cite{yfs},
are free of infrared singularities to all orders in $\alpha$. We use an obvious
notation for the 4-momenta $\{p_i\}$ for the scattering charged particles
$\{f_i\}$ and the infrared functions $B,~\tilde{B}$, and $D$ are as defined in Ref.~\cite{yfs}. The exactness of (\ref{subp15}) is essential for precision
theory applications.\par
The presentation is organized as follows. In the next Sections, we review
the applications of (\ref{subp15}). We discuss in this connection
the period before precision
electroweak(EW) physics at LEP/SLC, the era of precision EW physics,
the applications of the QCD extension of (\ref{subp15}) for precision 
LHC physics and recent
results obtained from applications of the extension of (\ref{subp15})
for quantum general relativity. We conclude with some discussion
of possible future applications. The Appendix gives an 
example of uncited impact of our calculations.\par
\section{Applications: Comparative Observations}
The original applications~\cite{yfs} of (\ref{subp15}) were
at the precision of the leading term, the $\bar\beta_0$-level, in which one 
retains only the $n=0$-term therein. The 4-momentum conservation
in (\ref{subp15}) is then treated exactly, which
necessitates integration over the y-dependent
exponential factors therein. This was done in Ref.~\cite{yfs} already,
with the result, for example, for initial state radiation(ISR) in 
$e^+e^-$ annihilation,
\begin{equation}
d\sigma_{exp}\cong\gamma F_{YFS}(\gamma)(1-z)^{\gamma-1}\sigma_Bdz
\label{app-b0}
\end{equation}      
where we have defined $z=s'/s$, $\gamma=\frac{2\alpha}{\pi}
(\ln\frac{s}{m^2}-1)$, and 
\begin{equation}
F_{YFS}(\gamma)=\frac{e^{-C\gamma}}{\Gamma(1+\gamma)}.
\end{equation} 
Here, $C=0.5772\ldots$ is Euler's constant and $\sigma_B$ is the respective
Born cross section. Only the leading terms in $\gamma$ are then retained
in this $\bar\beta_0$-level approximation~\cite{yfs}. The accuracy is
expected to be in the $\lesssim 10\%$ regime, which is quite adequate for
applications in which there were errors on $\sigma_B$
that could be much larger.
It is also important to note that these early applications of (\ref{subp15})
were (semi-)analytical in nature.\par
The LEP1/SLC, LEP2 era marked the application of (\ref{subp15}) to precision
predictions from quantum field theory via exact Monte Carlo methods.
The collaboration in this connection 
between the author and Staszek (Prof. Jadach) started
in the 1985-1986 time frame as a result of a Radiative Corrections
Workshop organized at SLAC 
by Prof. G. Feldman, who at that time was a Spokesman
for the MkII Collaboration at the SLC. We were both invited to participate
in that workshop and as a result we began discussion of the feasibility
to realize the exact result (\ref{subp15}) by Monte Carlo methods~\footnote{
This was a long and technical discussion, some of it done on walks
in the Tatra Mountains at a Zakopane Summer School, for example.}. 
The key issue, after much sucessful discussion on other issues, 
such as our reduction procedure~\cite{jw1prdyfs1}, etc., 
was the realization by Monte Carlo methods of the 
factor $e^D$ in (\ref{subp15}). The pioneering solution 
was given by Prof. Jadach in Ref.~\cite{sj1}. The title of the paper,
``Yennie-Frautschi-Suura Soft Photons in the Monte Carlo Event Generators'',
underscores how important it was to the Jadach-Ward approach to precision
theory for quantum field theory predictions for physical processes:
it opened the way to use the exact result (\ref{subp15}) 
via Monte Carlo methods so that arbitrarily precise 
predictions could be obtained on an event-by-event basis. 
The solution presented in Ref.~\cite{sj1} is to date the
only such solution known and thus is a true testament to the genius of its
creator.\par
With the complete set of ingredients now in place to realize (\ref{subp15}),
we published in 1988 in Ref.~\cite{jw1prdyfs1} the first realistic MC 
for precision SLC/LEP1 physics,
YFS1, an exact ${\cal O}(\alpha)$, YFS-exponentiated multiple photon
MC for $e^+e^-\rightarrow f\bar{f}+n(\gamma)$, ~$f\neq e$. Here, the modifier
``YFS'' denotes that the exponentiation is the 
resummation given by (\ref{subp15}). As we discuss in Ref.~\cite{jw1prdyfs1},
the precision tag for YFS1 in Z physics is $\lesssim 1\%$. This was followed in 1989 with the publication in Ref.~\cite{bhl1jw} of the first realistic
exact ${\cal O}(\alpha)$, YFS-exponentiated multiple photon
MC for $e^+e^-\rightarrow e^+e^-+n(\gamma)$ at low angles, BHLUMI1.0, 
for Z physics, where the primary applications were 
precision luminosity predictions. Again, the precision tag is $\lesssim 1\%$.
\par
The large number of Z's at LEP1 ($2\times 10^7$ were detected) 
necessitated per mille level theory precision in order that the 
theoretical error would not compromise the outstanding experimental 
error in the attendant tests of the EW and QCD theories. We therefore 
developed the YFS2 and YFS3 level MC realizations of
(\ref{subp15}) in Refs.~\cite{jwyfs2,jwyfs3}, wherein the precision tags are
0.1\% for initial state radiation and for the combination of 
initial state and final state radiation, respectively.\par
Continuing in this way, working as well with our collaborators M. Melles,
W. Placzek, E. Richter-Was, M. Skrzypek, Z. Was and S. Yost, we have 
developed the following YFS MC event generators, 
all realizations of (\ref{subp15}):
KORALZ3.8,4.04~\cite{krlz1} with 0.1\% precision tag on 2f production 
at the Z regime in LEP1/SLC; BHLUMI2.01,2.30,4.04~\cite{bhl2-4} 
for the LEP1/SLC luminosity 
process small angle Bhabha scattering with the final 
precision tag of 0.061\%(0l.054\%), according 
as one does not (does) implement the soft pairs effect from
either Ref.~\cite{bhl2.3,onicrsi}; and BHWIDE~\cite{bhw1} for the large angle
Bhabha scattering with precision tag 0.2\% at the Z regime at LEP1/SLC.\par
The advent of LEP2, and its attendant $2\times 10^5$ W pairs, created 
the need for precision predictions for W-pair productions and decay, 
the 4f background processes, radiative return Z production as 
well as the need for reliable 2Z production predictions. 
We developed~\cite{ceex1} the new coherent realization of (\ref{subp15})
to treat the Z-radiative return events at high precision by treating the 
real emission IR singularities at the level of amplitude in complete 
analogy with the original treatment of the virtual IR singularities 
by Yennie, Frautschi and Suura in Ref.~\cite{yfs}. We refer to this 
form of the theory as the CEEX theory. It is realized in the  
event generator KK MC~\cite{kkmc}, which gives 0.2\% precision on radiative 
return 2f production at LEP2 energies. In addition, for LEP2 our collaboration 
developed the MC's 
YFSWW3~\cite{yfsww3} with 0.4\% precision on WW production, 
KoralW(1.02,1.42)~\cite{krlw1} 
with 1.0\% precision on the 4f background processes,
KoralW1.51~\cite{krlw2}, the concurrent KoralW\&YFSWW3 MC, 
with 0.4\% on 4f production 
near the WW regime, and YFSZZ~\cite{yfszz} with 2\% precision 
for ZZ production. These are all state-of-the-art results for LEP2 based
on the rigorous MC realization of (\ref{subp15}) on an event-by-event basis.
We also determined~\cite{lep2yllw} the precisions of BHWIDE and 
BHLUMI at LEP2 as 0.4\% and 0.122\% respectively. We now present some
exemplary results based on these seminal calculations.\par
\subsection{Exemplary Results}
The MC KoralZ was a workhorse for LEP1,2 physics. As an example of its many applications, we illustrate with the analysis by the ALEPH Collaboration~\cite{alephkrlz} of their data on mu-pair production from 20 GeV to 136 GeV: We quote from
Ref.~\cite{alephkrlz},''In
order to study the effect of the experimental cuts, more than $2\times 10^6$
events were produced with full detector simulation, using the DYMU3[8] 
and KORALZ 4.0 [9] Monte Carlo event generators for the exclusive and inclusive
analysis, respectively, at several nominal LEP energies. Radiation of hard 
photons in the initial and final state is treated at ${\cal O}(\alpha)$ by
DYMU3 and at ${\cal O}(\alpha^2)$  by KORALZ 4.0. In KORALZ the radiation 
of soft photons is included at all orders by exponentiation.'' This is one 
of many examples.\par
\begin{figure}
\begin{center}
\includegraphics[width=100mm]{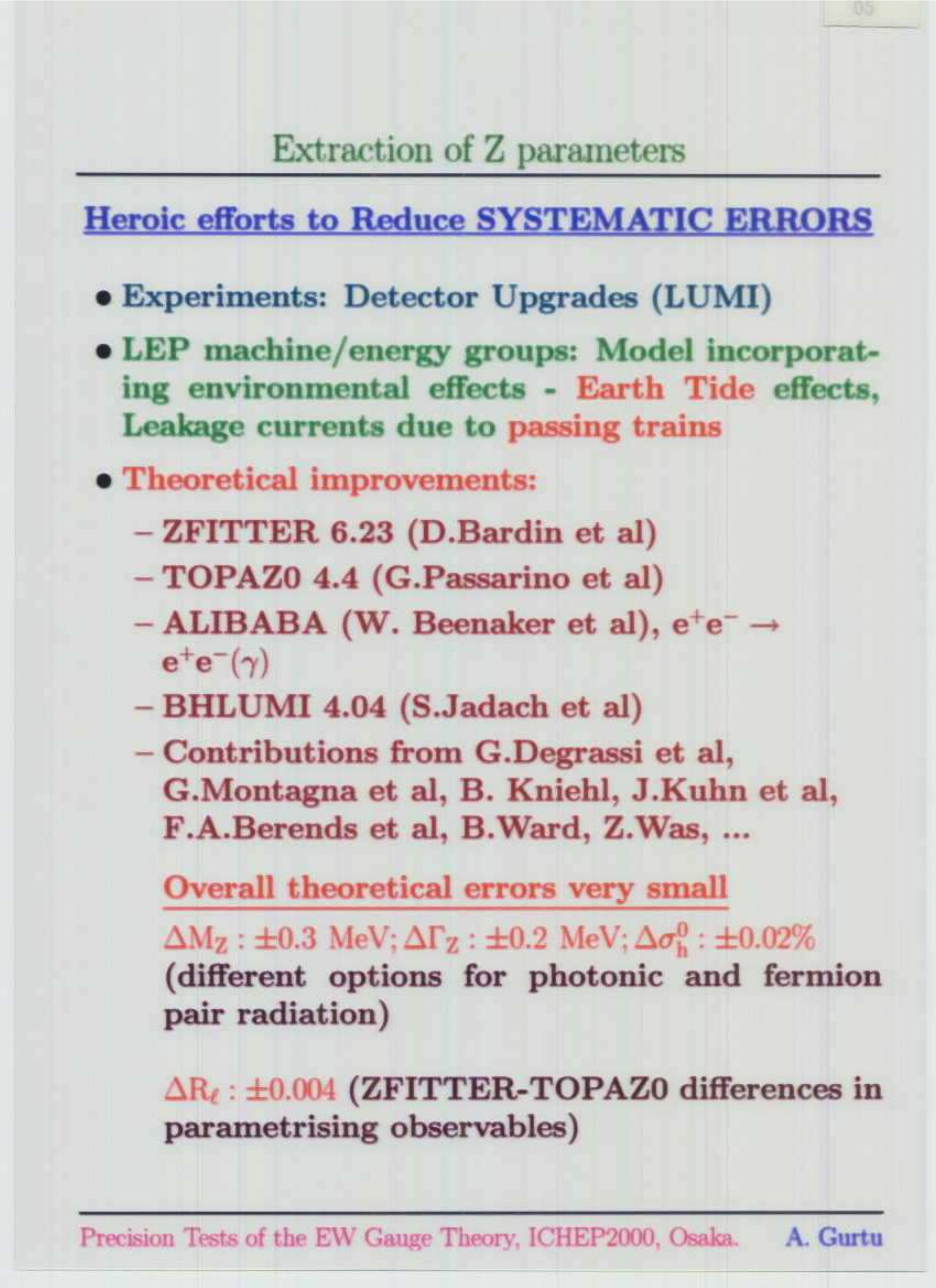}
\end{center}
\caption{\baselineskip=7mm  Summary of EW theory progress on Z physics as presented by Gurtu~\protect\cite{gurtu} in ICHEP2000.}
\label{fig1}
\end{figure}

 
In Fig.~\ref{fig1}, we show the summary of the progress on precision EW 
theory as presented by Gurtu in his review for ICHEP2000 at Osaka~\cite{gurtu}.
We see in the figure that he shows BHLUMI4.04 as a key element in these 
improvements which allowed the proper exploitation of the LEP data for 
precision SM tests.\par
\begin{figure}
\begin{center}
\includegraphics[width=100mm]{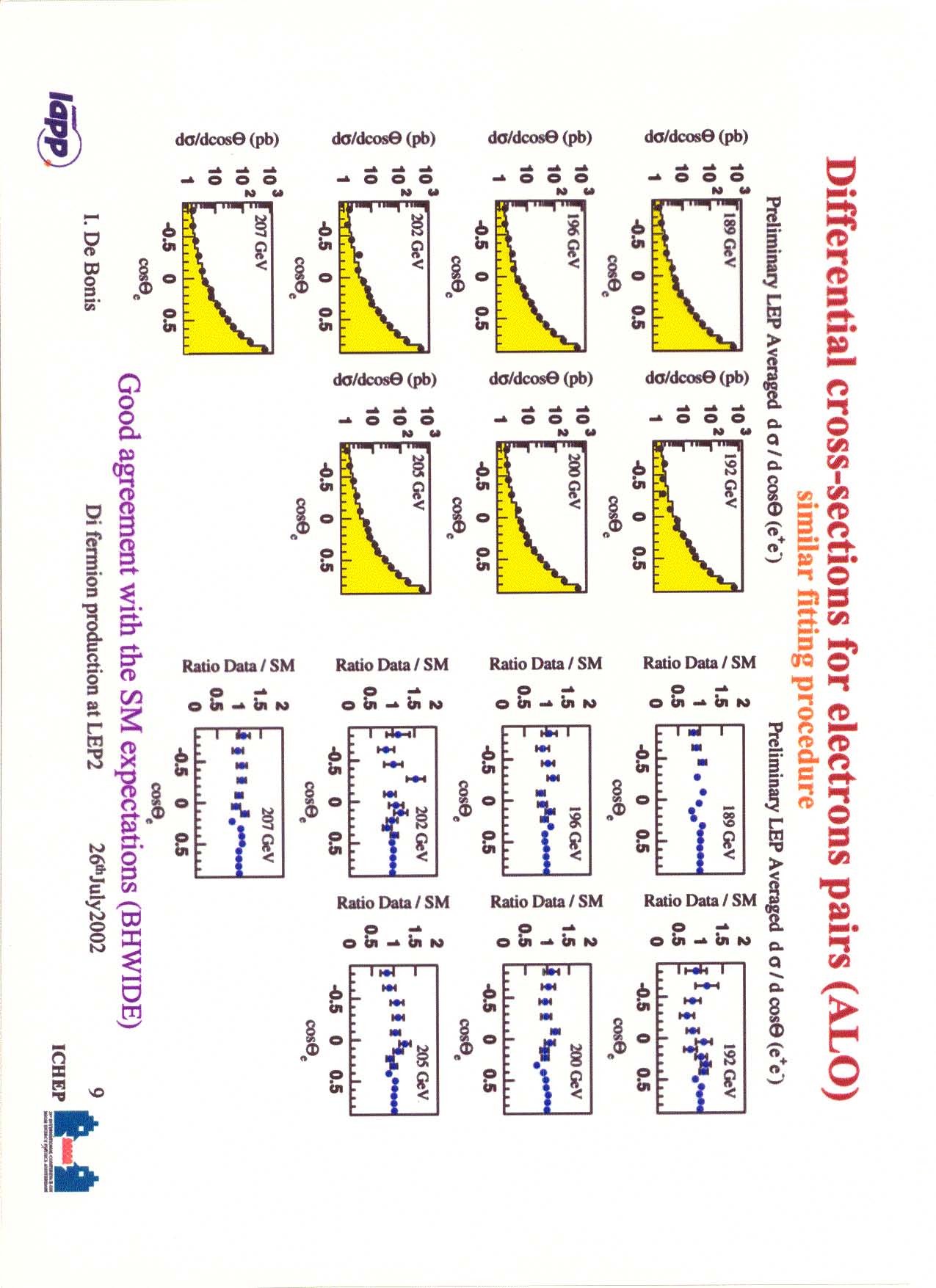}
\end{center}
\caption{\baselineskip=7mm  Comparison of BHWIDE with precision LEP data
as presented in Ref.~\cite{debonis1} at ICHEP2002.}
\label{fig2}
\end{figure}

For BHWIDE, there are also many examples of its seminal role in 
establishing the precision comparison between the Standard Model 
EW theory and the LEP data. We show in Fig.~\ref{fig2} the results presented
by De Bonis~\cite{debonis1} at ICHEP02, where he shows that BHWIDE gives
outstanding agreement with the LEP observations of large angle Bhabha 
scattering\footnote{The actual impact of BHWIDE on $e^+e^-$ annihilation
discovery physics is clouded by the exchange of e-mails with Drs. Marsiske and
MacFarlane shown in
the Appendix. Their Babar Collaboration have used 
the MC extensively as described by Dr. Marsiske but have
not referenced this use in their published papers, only in internal notes
as he describes. Such notes are not available to the public so we have no idea
as to what the actual impact of the calculation really has been.}.
\par
\begin{figure}
\begin{center}
\includegraphics[width=100mm]{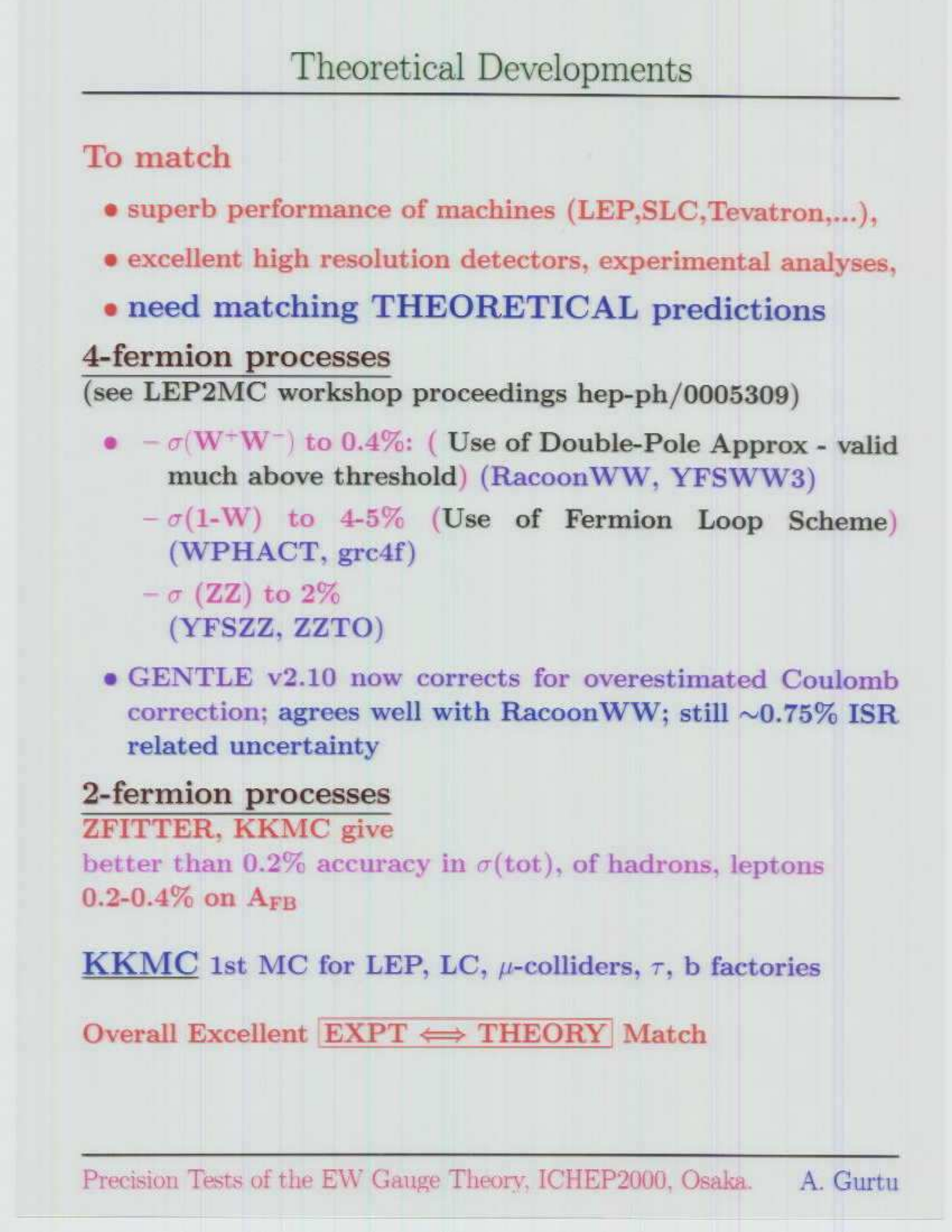}
\end{center}
\caption{\baselineskip=7mm  Comparison of YFSWW3 and RacoonWW with precision
LEP2 data as presented in Ref.~\protect\cite{gurtu} at ICHEP2000.}
\label{fig3}
\end{figure}
\begin{figure}
\begin{center}
\includegraphics[width=100mm]{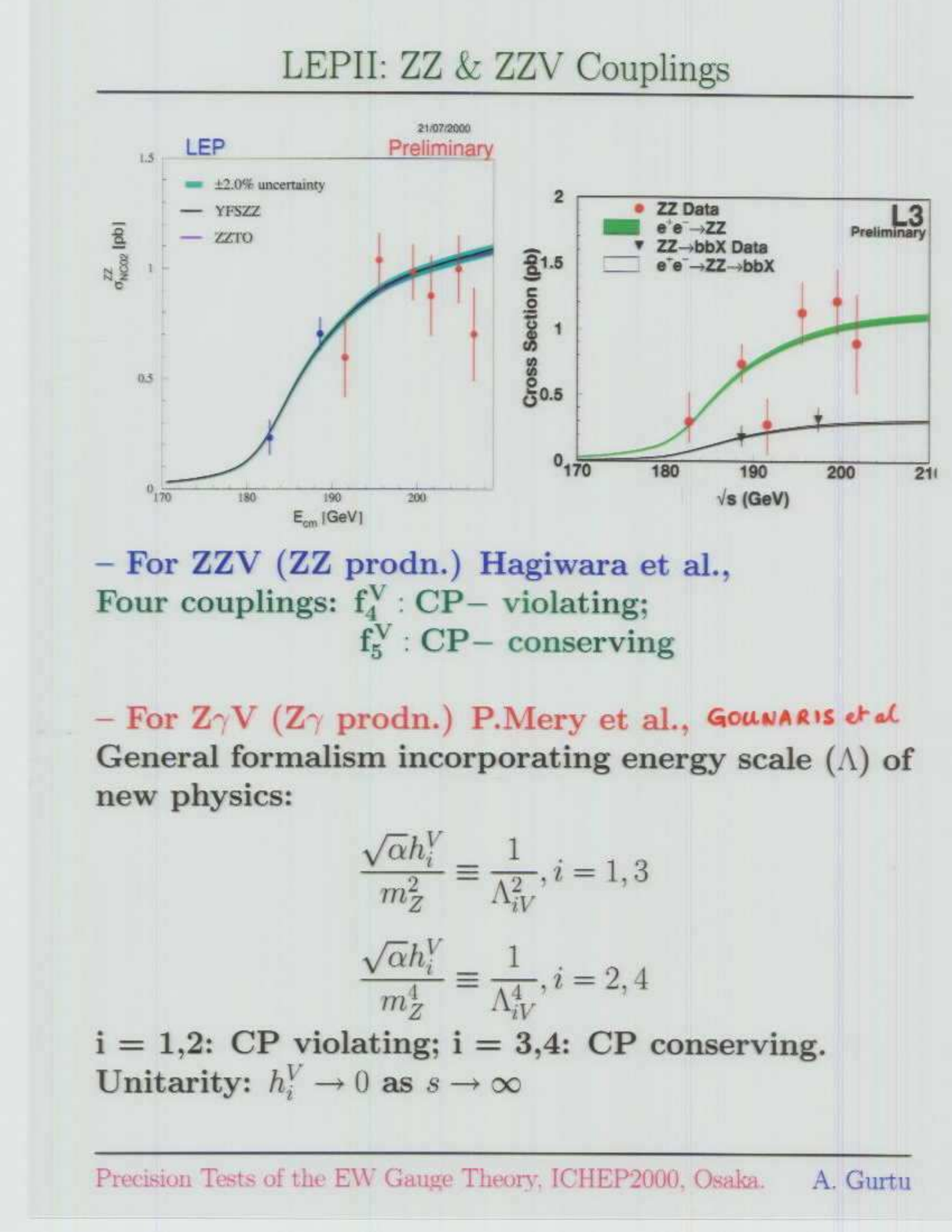}
\end{center}
\caption{\baselineskip=7mm  Comparison of YFSZZ with 
LEP2 Z-pair production data as presented in Ref.~\protect\cite{gurtu} 
at ICHEP2000.}
\label{fig4}
\end{figure}
\begin{figure}
\begin{center}
\includegraphics[width=100mm]{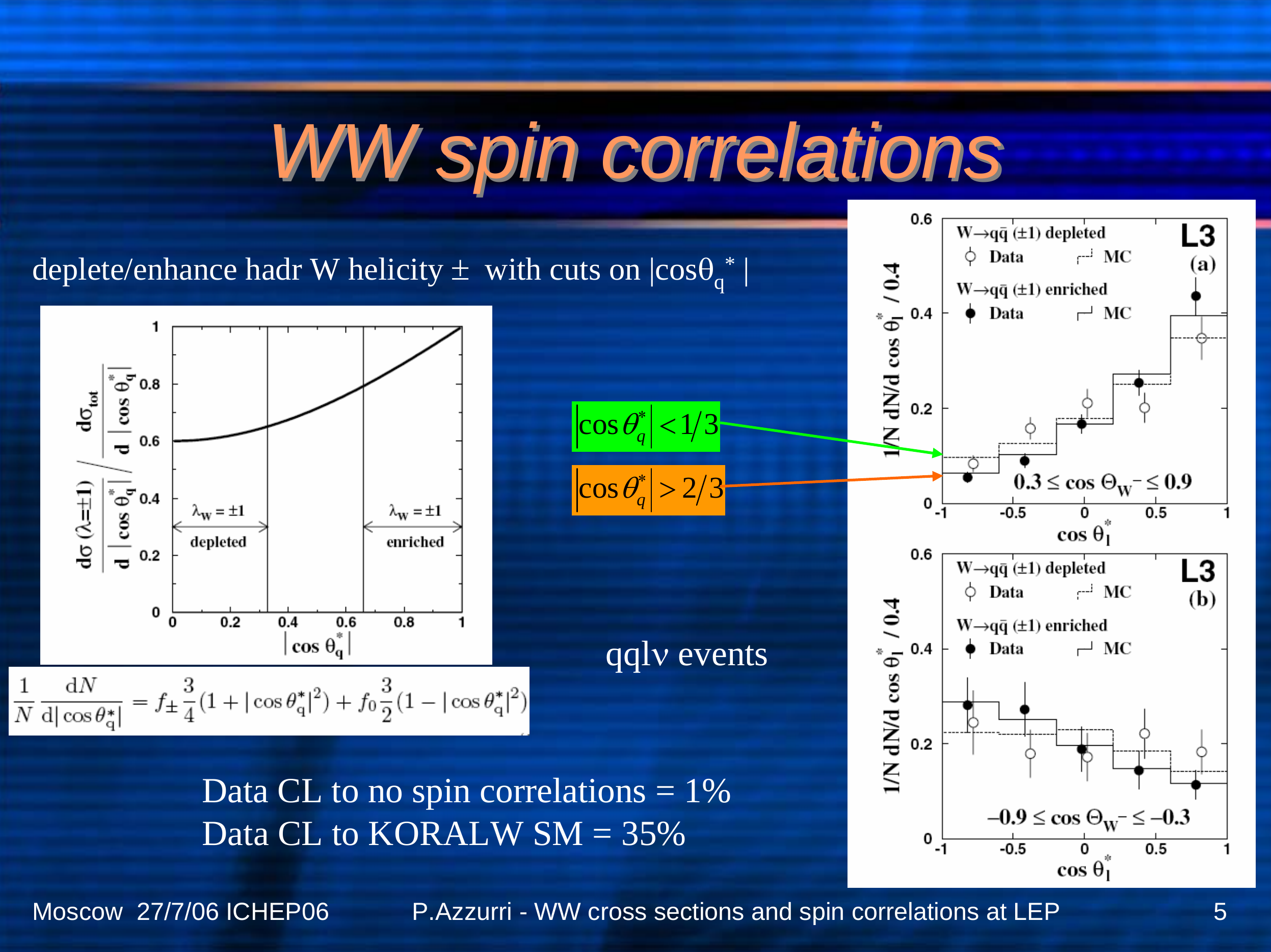}
\end{center}
\caption{\baselineskip=7mm  Comparison of KoralW with 
LEP2 WW/4f spin correlation production data as presented in Ref.~\cite{azzuri} 
at ICHEP2006.}
\label{fig5}
\end{figure}

For YFSWW and KK MC, there are also many examples of their seminal role in
precision LEP physics. To illustrate, we use again an example for from Ref.~\cite{gurtu} as shown in Fig.~\ref{fig3} which summarizes the progress
in theory for 2f and 4f processes at LEP1,2 for ICHEP2000. The MC YFSZZ
is also featured in Fig.~\ref{fig3}, as it provided state-of-the-art
simulations for the Z-pair production data at LEP2. We see then in 
Figs.~\ref{fig4},~\ref{fig5} that the YFSWW3, along with RacoonWW~\cite{racoonww}, did indeed 
establish the proper normalization and simulation of the LEP2 WW pair production as predicted by the 't Hooft-Veltman non-Abelian gauge theory renormalization
theory~\cite{tHftvelt} and that YFSZZ did indeed provide state-of-the-art
Z-pair production simulation for the LEP2 data.
\par
The Monte Carlo KoralW has played an essential role in the 4f/WW data analysis
as well, providing as it did, precision simulation of the background processes
for W-pairs as we have indicated. This is illustrated in Fig.~\ref{fig5}.
What we have illustrated are examples that indicate the broad effect
that the Monte Carlo realization of (\ref{subp15}) has had on tests
of the SM using precision LEP data.\par
\begin{figure}
\begin{center}
\includegraphics[width=100mm]{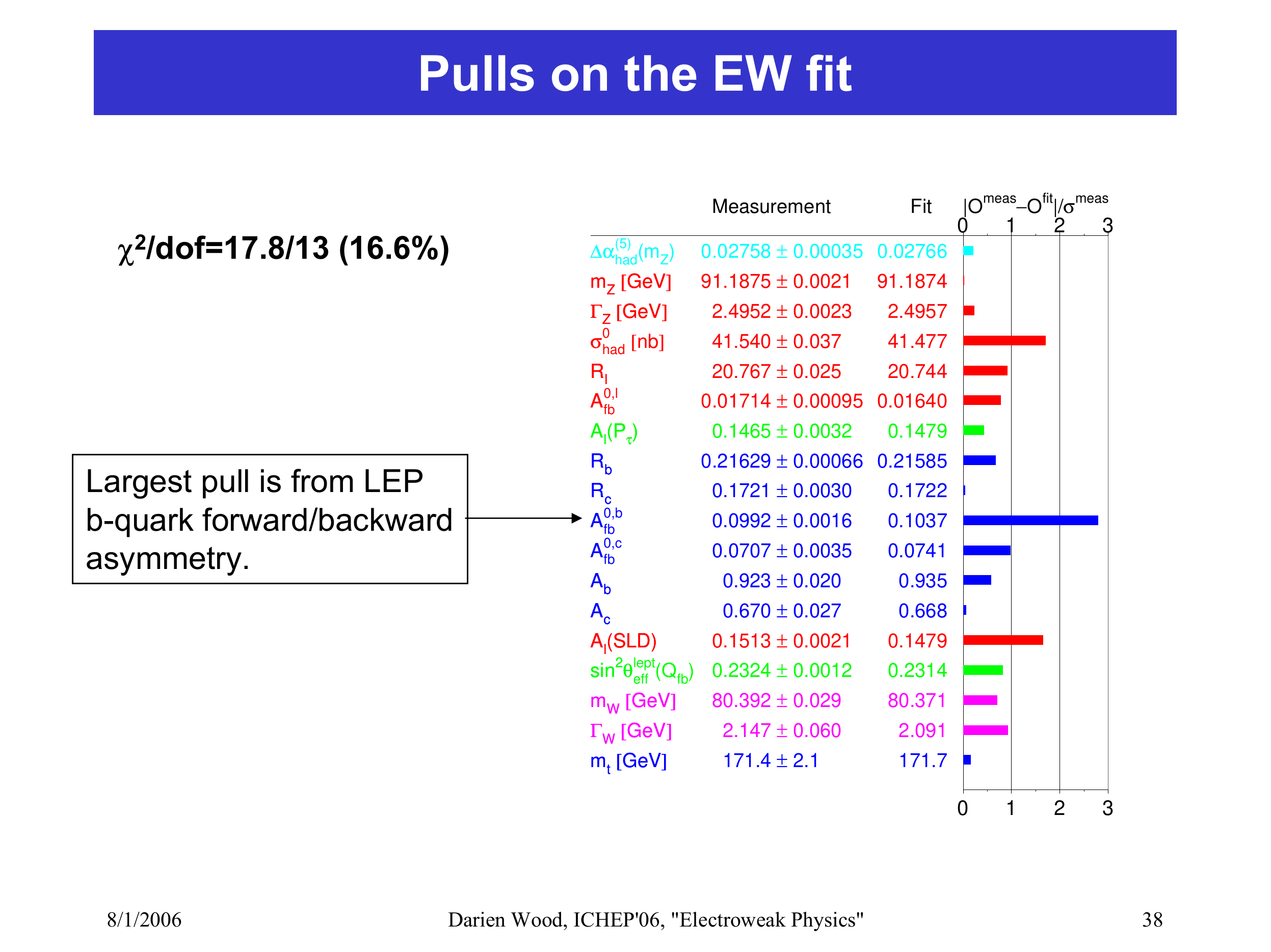}
\end{center}
\caption{\baselineskip=7mm  Comparison of precision EW data with 
the SM theory as presented in Ref.~\cite{dwood} 
at ICHEP2006.}
\label{fig6}
\end{figure}
\begin{figure}
\begin{center}
\includegraphics[width=100mm]{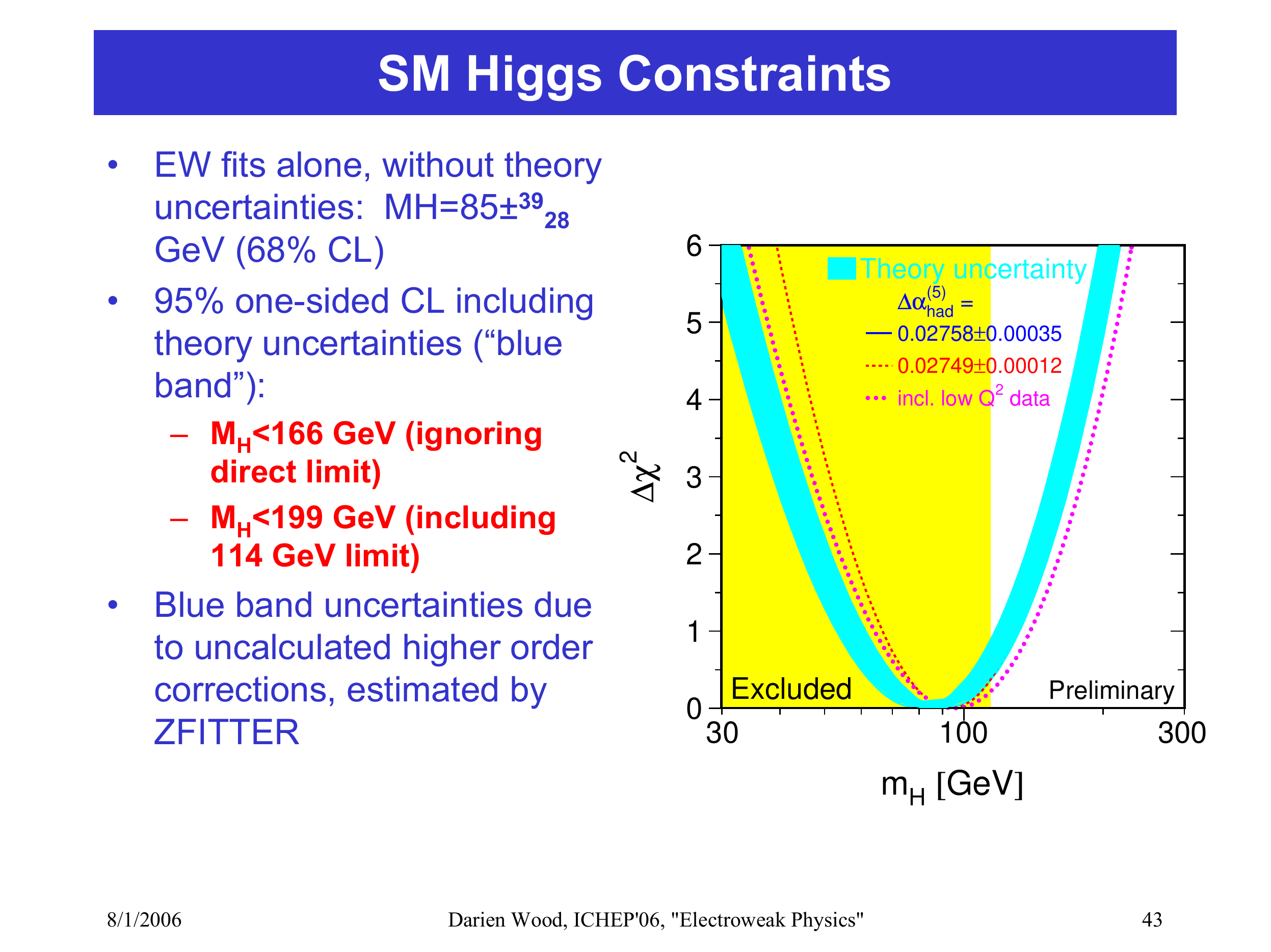}
\end{center}
\caption{\baselineskip=7mm  Implications for the mass of the SM
Higgs particle from the SM EW fit to precision LEP data
as presented in Ref.~\cite{dwood} 
at ICHEP2006.}
\label{fig7}
\end{figure}

Indeed, these precision calculations, which we need to emphasize employed
as well the pioneering EW libraries of Refs.~\cite{ewlib} in isolating
some of the purely weak exact results in the residuals $\bar\beta_n$,
have played essential roles in determining the degree of agreement
between then SM non-Abelian loop corrections to precision observables
and the value of these effects as measured by LEP data. This is illustrated
in Fig.~\ref{fig6} as it is presented in Ref.~\cite{dwood} at ICHEP06. The many consequences of the latter comparison, such as its implications
for the mass of the still-sought SM Higgs particle -- a main objective
for discovery at LHC, are illustrated in
Fig.~\ref{fig7}. The precision comparison between the SM expectations
and the LEP data establish the correctness of the 't Hooft-Veltman renormalization theory for non-Abelian gauge theories at the one-loop level and give
us confidence that the origin of EW symmetry breaking, as it is represented
by the Higgs boson, is within reach of LHC experimentation. In addition,
when the precise value of the running $\alpha_s(Q)$ is extracted for the
the LEP data and compared with data at lower energies~\cite{siggi}, one 
also obtains experimental proof of the running of the latter coupling as
predicted by the asymptotic freedom discovery of 
Gross, Wilczek~\cite{gwilzk} and Politzer~\cite{poltzr}. The Royal
Swedish Academy~\cite{royswedacad} has emphasized 
these points in awarding the 1999 Nobel Prize
in Physics to Profs. G. 't Hooft and M. Veltman, with the citation''...for elucidating the quantum nature of the electroweak interactions in physics...The theory's predictions verified...large quantities of of W and Z have recently been produced under controlled conditions at the LEP accelerator at CERN. Comparisons between measurements and calculations have all the time showed great agreement, thus supporting the theory's predictions...'',
and the 2004 Nobel Prize in Physics to Profs. D.J. Gross, F. Wilczek and H.D. Politzer, with the citation''...The theory has been tested in great detail, in
particular during recent years at the European Laboratory for Particle Physics, CERN, in Geneva...''. Prof. Jadach and his collaborators have made via
YFS-based MC methods an essential contribution to the realization of the
two respectively cited precision studies.\par
\section{QCD and QED$\otimes$QCD Extension}
Already at the start of the preparations for the physics program 
for the now canceled SSC, we moved our attention to the application 
of the analog of (\ref{subp15})
to the QCD theory in Refs.~\cite{delaney}. This development has resulted
in the QCD resummation formula~\cite{qcdexp}, for the processes
$f_1(p_1)+f_2(q_1)\rightarrow f_3(p_2)+f_4(q_2)+n(G)$,
\begin{equation}
\begin{split}
d\hat\sigma_{\rm exp}
         &=e^{\rm \Sigma_{IR}(QCD)}\sum_{n=0}^\infty\frac{1}{n!}\int\prod_{j=1}^n{d^3
k_j\over k_j}\int{d^4y\over(2\pi)^4}e^{iy\cdot(p_1+q_1-p_2-q_2-\sum k_j)+
D_\rQCD}\\
&*\tilde{\bar\beta}_n(k_1,\ldots,k_n){d^3p_2\over p_2^{\,0}}{d^3q_2\over
q_2^{\,0}}
\end{split}
\label{subp15qcd}
\end{equation}
where now the hard gluon residuals 
$\tilde{\bar\beta}_n(k_1,\ldots,k_n)$
are free of all infrared divergences to all 
orders in $\alpha_s(Q)$. The functions
$SUM_{IR}(QCD), D_\rQCD$, together with 
the attendant basic infrared functions 
$B^{nls}_{QCD},{\tilde B}^{nls}_{QCD},{\tilde S}^{nls}_{QCD}$ 
are specified in Ref.~\cite{qcdexp}. Here, $Q$ is the relevant hard scale.
We have shown that (\ref{subp15qcd})
leads to an independent cross check of the size of threshold resummation
effects in $t\bar{t}$ production at FNAL at the 1\% level as found in Ref.~\cite{catman}. More recently, realizing that for LHC physics the EW corrections
can be significant in a 1\% error budget, we have extended the result
(\ref{subp15qcd}) to the simultaneous resummation of QED and QCD,
QED$\otimes$QCD resummation~\cite{qcedexp}, 
\begin{equation}
\begin{split}
d\hat{\sigma}_{\rm exp} &= e^{\rm SUM_{IR}(QCED)}\cr
   &\sum_{{n,m}=0}^\infty\frac{1}{n!m!}\int\prod_{j_1=1}^n\frac{d^3k_{j_1}}{k_{j_1}} 
\prod_{j_2=1}^m\frac{d^3{k'}_{j_2}}{{k'}_{j_2}}
\int\frac{d^4y}{(2\pi)^4}\cr
&e^{iy\cdot(p_1+q_1-p_2-q_2-\sum k_{j_1}-\sum {k'}_{j_2})+
D_{\rm QCED}} \cr
&\tilde{\bar\beta}_{n,m}(k_1,\ldots,k_n;k'_1,\ldots,k'_m)\frac{d^3p_2}{p_2^{\,0}}\frac{d^3q_2}{q_2^{\,0}},
\end{split}
\label{qced}
\end{equation}
\noindent
where the new YFS~\cite{yfs,jw1prdyfs1} 
residuals, defined in Ref.~\cite{qcedexp}, 
$\tilde{\bar\beta}_{n,m}(k_1,\ldots,k_n;k'_1,\ldots,k'_m)$, with $n$ hard gluons and $m$ hard photons,
represent the successive application
of the YFS expansion first for QCD and subsequently for QED. The
functions ${\rm SUM_{IR}(QCED)},D_{\rm QCED}$ are determined
from their analogs ${\rm SUM_{IR}(QCD)},D_\rQCD$ via the
substitutions
{\small
\begin{eqnarray}
B^{nls}_{QCD} \rightarrow B^{nls}_{QCD}+B^{nls}_{QED}\equiv B^{nls}_{QCED},\cr
{\tilde B}^{nls}_{QCD}\rightarrow {\tilde B}^{nls}_{QCD}+{\tilde B}^{nls}_{QED}\equiv {\tilde B}^{nls}_{QCED}, \cr
{\tilde S}^{nls}_{QCD}\rightarrow {\tilde S}^{nls}_{QCD}+{\tilde S}^{nls}_{QED}\equiv {\tilde S}^{nls}_{QCED}
\label{irsub}
\end{eqnarray}}  
everywhere in expressions for the
latter functions given in Refs.~\cite{qcdexp}.
The residuals $\tilde{\bar\beta}_{n,m}(k_1,\ldots,k_n;k'_1,\ldots,k'_m)$ 
are free of all infrared singularities.
The result in (\ref{qced}) is a representation that is exact
and that can therefore be used to make contact with parton shower 
MC's without double counting or the unnecessary averaging of effects
such as the gluon azimuthal angular distribution relative to its
parent's momentum direction.\par
Indeed, from the result (\ref{qced}) and the standard formula
for the hadron cross section,
\begin{equation}
d\sigma=\sum_{i,j}\int dx_1dx_2F_i(x_1)F_j(x_2)d\hat\sigma_{\rm exp}
\end{equation}
we have immediately two issues to address: shower/ME matching, which we do preferably by shower-subtracted residuals, $\tilde{\bar\beta}_{m,n} \rightarrow \hat{\tilde{\bar\beta}}_{m,n}$,
as presented in Ref.~\cite{hera-lhc}, and for MC stability, IR-improved
DGLAP-CS theory~\cite{irdglap}, a new exponentiated scheme for the respective
kernels, $P_{AB}$,
reduced cross sections, and parton distributions, 
\begin{equation}
\begin{split}
F_1,\;\hat{\sigma}&\rightarrow F_i',\;\hat{\sigma}'\quad\text{for}\\
P_{qq}&\rightarrow P_{qq}^{exp}=C_FF_{YFS}(\gamma_q)e^{\frac{1}{2}\delta_q}\frac{1+z^2}{1-z}(1-z)^{\gamma_q}, \; \text{,etc.,}
\end{split}
\end{equation}
giving the same value for the respective hadron cross section $\sigma$,
with improved MC stability.\par
In addition, other technical checks are now open, such as
the issue of setting all quark masses $m_q$ to zero in the ISR at
${\cal O}(\alpha_s^n),\; n\ge 2$ due to the 
theorem in Refs.~\cite{dieleto, cat1},
according to which there is a lack of Bloch-Nordsieck
cancellation of IR singularities unless $m_q=0$. We show in Ref.~\cite{nbln1}
that the result (\ref{subp15qcd}) obviates this theorem.\par
The matter of an independent cross-check of the standard backward evolution
algorithm for the parton shower itself~\cite{sjos1} is also
under study with the results of Refs.~\cite{showers1,showers2}. 
Staszek's group
are actively involved in this development.\par
There are many more issues which we do not have space to list here: They are
all under study. All of the necessary theoretical formalism is at hand --
this underscores the need to support exact results for higher order
calculations, cross checks, tests, etc., to prove 1\% precision for LHC
luminosity processes for example. We can not emphasize this too much.\par
\section{Extension to QGR}
The exactness of the re-arrangement means that we can apply the same
resummation algebra to quantum gravity~\cite{bw1,bw2,bw3,bw4}. We find that the
scalar propagator for mass $m$ resums in quantum gravity to
\begin{equation}
i\Delta'_F(k)|_{\text{resummed}} =  \frac{ie^{B''_g(k)}}{(k^2-m^2-\Sigma'_s+i\epsilon)}
\label{resum}
\end{equation}
for{\small ~~~($\Delta =k^2 - m^2$)
\begin{equation}
\begin{split} 
B''_g(k)&= -2i\kappa^2k^4\frac{\int d^4\ell}{16\pi^4}\frac{1}{\ell^2-\lambda^2+i\epsilon}\\
&\qquad\frac{1}{(\ell^2+2\ell k+\Delta +i\epsilon)^2}\\
&=\frac{\kappa^2|k^2|}{8\pi^2}\ln\left(\frac{m^2}{m^2+|k^2|}\right),       
\end{split}
\label{yfs1} 
\end{equation}}
where the latter form holds for the UV regime, so that (\ref{resum}) 
falls faster than any power of $|k^2|$. An analogous result~\cite{bw1} holds
for m=0. We also note that, as $\Sigma'_s$ starts in ${\cal O}(\kappa^2)$,
we may drop it in calculating one-loop effects. It follows that
when the respective analogs of (\ref{resum}) are used, one-loop 
corrections are finite. In fact, it can be shown that the use of
our resummed propagators renders all quantum 
gravity loops UV finite~\cite{bw1,bw2,bw3,bw4}. We have called this representation
of the quantum theory of general relativity resummed quantum gravity (RQG).
Its phenomenology is under study: we show in Refs.~\cite{bw4} that the final
state of Hawking radiation~\cite{hawk1} leads to Planck scale cosmic rays, etc.\par
\section{Future}
All of the developments extend to higher energy and/or higher precision
at lower energies down to 1GeV: at the B-Factory, the KK MC is already
in wide use~\cite{bfact}; at the $\Phi$-factories there are cross checks ~\cite{crossck} using KK MC with the distributions of the
program PHOKHARA ~\cite{phokra}, etc.\par
For higher energies in $e^+e^-$ annihilation, YFSWW, KoralW, BHWIDE, BHLUMI and
KK MC are all in play. For example, the ILC luminosity requirement 
~\cite{brau} is 0.01\%. 
\begin{figure}
\begin{center}
\includegraphics[width=190mm,height=340mm]{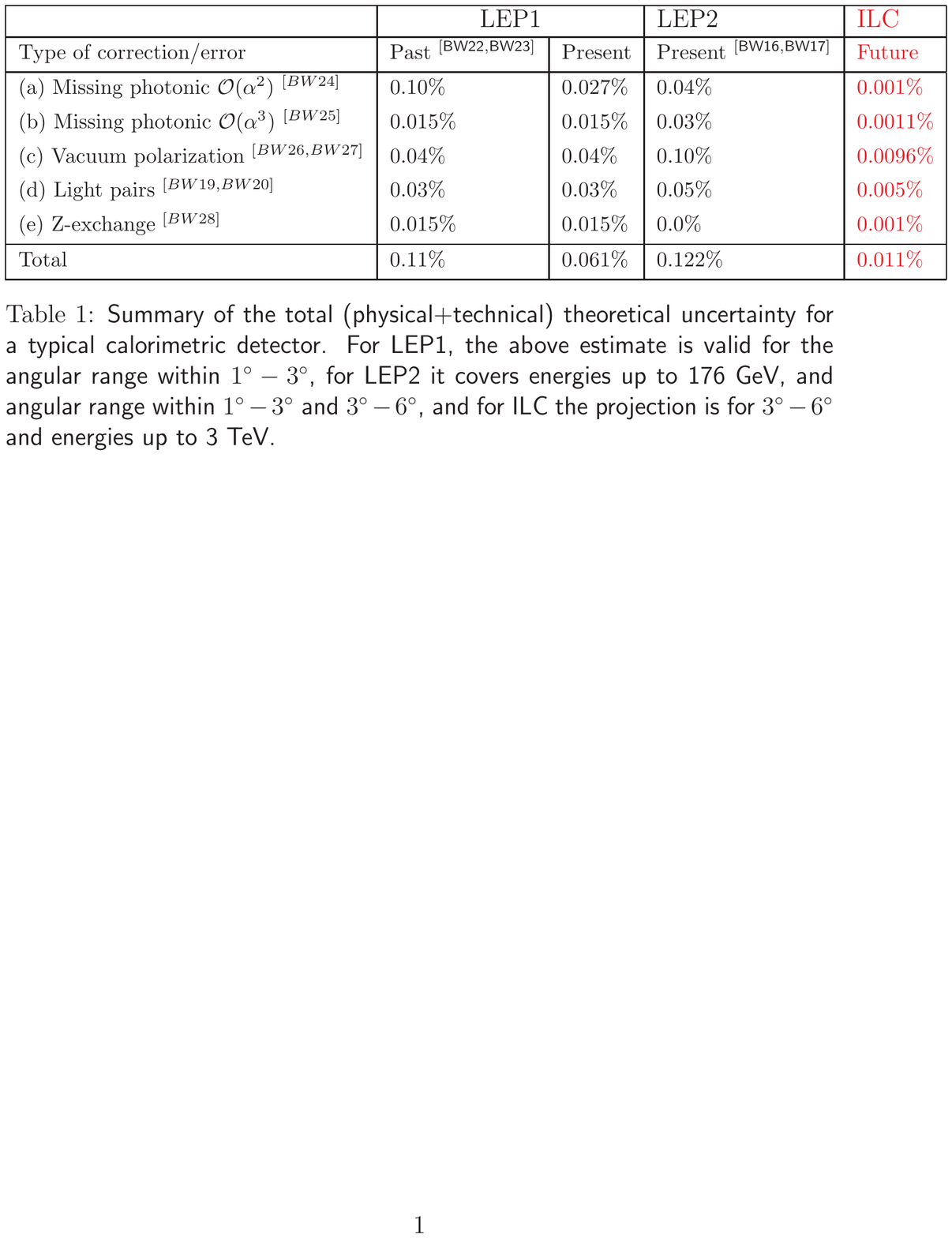}
\end{center}
\label{fig8}
\end{figure} 
We show in Table 1 what the extension of BHLUMI
from version 4.04 to version 5.0 for 0.011\% would involve(The references
in the table can be found in Ref.~\cite{doe2007}). We have
already explained in Ref.~\cite{doe2007} what this achievement would 
involve and how long in time it would take, about 3 years. Again,
it is all a question of support. It may be needed by 2025-2030?\par
From 1987 to 2027, what fun it is! And, we all owe a debt of special thanks
to Staszek for his seminal role in it.\par  
\section*{Appendix: Example of Internal Un-cited Use of BHWIDE}
In this appendix we record an email exchange we have had with members of the
BaBar Collaboration regarding the un-cited use of BHWIDE. From the exchange,
one can see that BHWIDE was used extensively by the collaboration 
without ever being referenced in whatever published papers were 
produced with the aid of its use. Even in the paper in the 
Nucl. Inst. and Methods journal on the detector itself`\cite{nimbabar}, 
BHWIDE was not referenced for the simulation of wide angle 
Bhabha's: was some other calculation used? We will never know.\par
==E-mail: Drs. Marsiske and MacFarlane(Spokesman) of BaBar and the author ==\\

            -----Original Message-----

From: Ward, B.F.L.

Sent: Tuesday, February 01, 2005 7:43 AM

To: 'dbmacf@slac.stanford.edu'

Subject: RE: RE: BHWIDE

Hello David,

   Thanks again in advance.

   Best regards,

   Bennie

Bennie F.L. Ward,

Distinguished Professor and Chairman,

Department of Physics,

Baylor University,

P.O. Box 97316

Waco, TX 76798-7316

Tel. 254-710-4878, Fax 254-710-3878

-----Original Message-----

From: David B. MacFarlane [mailto:dbmacf@slac.stanford.edu]

Sent: Monday, January 31, 2005 11:29 PM

To: Ward, B.F.L.; Staszek.Jadach@cern.ch; Wieslaw.Placzek@cern.ch

Subject: RE: RE: BHWIDE

Bernie:

Thanks for bringing this to my attention. I will have to look into the matter, which will take a little time, but I hope to get back to you by week's end.

Regards,

David

$>$ -----Original Message-----

$>$ From: Ward, B.F.L. [mailto:BFL\_Ward@baylor.edu]

$>$ Sent: Monday, January 31, 2005 10:58 AM

$>$ To: dbmacf@slac.stanford.edu; Staszek.Jadach@cern.ch;

$>$ Wieslaw.Placzek@cern.ch

$>$ Subject: FW: RE: BHWIDE

$>$

$>$ Hello David,

$>$     As you can see form(sic) my communications with Helmut Marsiske below,

$>$ our calculation BHWIDE, which realizes YFS exponentiated exact

$>$ O(alpha) multiple photon radiative effects on an event-by-event

$>$ basis by MC methods, was introduced into BaBar by Helmut with our

$>$ assistance several years ago.

$>$ He explains below that the program " BHWIDE has since been used

$>$ *extensively* at BABAR and is

$>$ *crucial* for our physics output: it is *the* generator to create MC

$>$ samples of (mostly) non-radiative as well as radiative Bhabhas, and to

$>$ calculate the necessary cross sections and efficiencies. The

$>$ non-radiative Bhabhas are used for our luminosity measurement and for

$>$ the single-crystal calibration of the electromagnetic calorimeter. The

$>$ radiative Bhabhas are used for our cluster energy calibration and for

$>$ E/p studies in connection with electron identification.

$>$ BHWIDE has been referenced (i.e., Phys. Lett. B 390 (1997) 298) in numerous *internal*

$>$ documents dealing with the above-mentioned areas of luminosity,

$>$ calibration, and PID. I'm not sure, though, whether it has

$>$ made it into any of our journal publications."

$>$     According to the latest SPIRES data that I have, the program has

$>$ never been referenced by BaBar's many published papers. This is very

$>$ hard on its authors for the obvious reasons: promotions,

$>$ funding awards, etc., in our field, as you well 
$>$ know as a Collaboration Spokesman, are

$>$ all ultimately very much dependent on ones citations, especially

$>$ citations by a flagship experiment such as yours.

$>$ Thus, I am writing to ask you why a calculation which has apparently

$>$ been very helpful in your

$>$ physics analysis has never been cited as having played any such role

$>$ therein in the published

$>$ literature? Mentioning BHWIDE in your private Collaboration notes as

$>$ Helmut indiactes(sic) does not really give its authors their proper credit,

$>$ as these notes are not read by the general peer-reviewing public.

$>$     Thanks in advance.

$>$     Best regards,

$>$     Bennie
$>$ Bennie F.L. Ward,

$>$ Distinguished Professor and Chairman,

$>$ Department of Physics,

$>$ Baylor University,

$>$ P.O. Box 97316

$>$ Waco, TX 76798-7316

$>$ Tel. 254-710-4878, Fax 254-710-3878

Date: Mon, 22 Apr 2002 14:23:51 -0400

From: bflward $<$bflward@utk.edu$>$

To: Stanislaw.Jadach@cern.ch, Wiesiek.Placzek@cern.ch

Cc: bflward@utk.edu

Subject: FWD: RE: BHWIDE

Hello Staszek and Wiesiek,

   He says they are going to do better? We will see.

   Thanks.

   Best regards,

   Bennie

$>$===== Original Message From Helmut Marsiske

$>$$<$marsiske@SLAC.Stanford.EDU$>$                              

=====

Bennie-

of course there is no policy in BABAR against referencing your BHWIDE, or any other, paper, and as I said: it has been referenced in internal notes. The fact that it wasn't mentioned in the NIM detector paper must have been a plain oversight. Sorry for that. We should try to do better in future papers...

   -H-

------------------------------------------------------------------------

|                                                                      |

| Dr. Helmut Marsiske          Stanford Linear Accelerator Center      |

| Stanford University                                                  |

| SLAC, Mail Stop 95           E-mail: MARSISKE@SLAC.Stanford.edu      |

| 2575 Sand Hill Road          Phone:  650-926-4333                    |

| Menlo Park, CA 94025         Fax:    650-926-2657                    |

| USA                          URL:    www.slac.stanford.edu/~marsiske |

|                                                                      |

------------------------------------------------------------------------

On Wed, 17 Apr 2002, bflward wrote:

$>$ Hello Helmut,

$>$    It is great to hear that BHWIDE has been useful to BaBar. What

$>$ would really help us is the referencing of the program 
$>$when it is used in your

$>$ preprints and publications, if this is possible -- at LEP, it is

$>$ routinely done and we have 84 citations in LEP publications. We seem

$>$ to have none in BaBar's?

$>$ For example, in your paper on the BaBar detector, hep-ex/0105044, on

$>$ page 10, you say you compare with the MC generator but you do not

$>$ reference which generator it is. If that was BHWIDE, then it would

$>$ really have helped us with our funding agencies, scientific

$>$ evaluations, etc. if you could have given us that reference in the

$>$ paper. Or, is there a policy in BaBar against this?

$>$     Thanks in advance.

$>$     Best regards,

$>$     Staszek Jadach, Wiesiek Placzek and Bennie

$>$ -----Original Message-----

$>$ $>$===== Original Message From Helmut Marsiske

$>$ $>$$<$marsiske@SLAC.Stanford.EDU$>$

$>$ =====

$>$ Hi Bennie,

$>$

$>$ indeed, BHWIDE has since been used *extensively* at BABAR and is

$>$ *crucial* for our physics output: it is *the* generator to create MC

$>$ samples of (mostly) non-radiative as well as radiative Bhabhas, and to

$>$ calculate the necessary cross sections and efficiencies. The

$>$ non-radiative Bhabhas are used for our luminosity measurement and for

$>$ the single-crystal calibration of the electromagnetic calorimeter. The

$>$ radiative Bhabhas are used for our cluster energy calibration and for

$>$ E/p studies in connection with electron identification. BHWIDE has

$>$ been referenced (i.e., Phys. Lett. B 390 (1997) 298) in numerous

$>$ *internal* documents dealing with the above-mentioned areas of

$>$ luminosity, calibration, and PID. I'm not sure, though, whether it has

$>$ made it into any of our journal publications.

$>$

$>$ Hope this helps,

$>$

$>$    Helmut

$>$

$>$ --------------------------------------------------------------

$>$ ----------

$>$ |                                                            

$>$          |

$>$ | Dr. Helmut Marsiske          Stanford Linear Accelerator

$>$ Center      |

$>$ | Stanford University                                        

$>$          |

$>$ | SLAC, Mail Stop 95           E-mail:

$>$ MARSISKE@SLAC.Stanford.edu      |

$>$ | 2575 Sand Hill Road          Phone:  650-926-4333          

$>$          |

$>$ | Menlo Park, CA 94025         Fax:    650-926-2657          

$>$          |

$>$ | USA                          URL:   

$>$ www.slac.stanford.edu/~marsiske |

$>$ |                                                            

$>$ --------------------------------------------------------------

$>$

$>$ On Mon, 8 Apr 2002, bflward wrote:

$>$

$>$ $>$ Hello Helmut,

$>$ $>$    If I recall correctly, you introduced BHWIDE into the BaBar

$>$ $>$ software? Has it actually been used for any analysis of wide angle

$>$ $>$ Bhabha's, etc., yet, and, if so, was that use referenced

$>$ anywhere in Babar preprints or publications? 

$>$I am having to explain my  existence

$>$ $>$ to my program manager for DoE ( i.e., he is cutting my

$>$ grant ) and any information like this would be very helpful, indeed.

$>$ $>$    Thanks in advance.

$>$ $>$    Best regards,

$>$ $>$    Bennie

\end{document}